\documentclass[twocolumn,amsmath,amssymb,prl,superscriptaddress,nobibnotes,showpacs]{revtex4}
\usepackage{graphicx}

\begin{document}
\title{Minimum Disturbance Measurement without Post-Selection}

\author{So-Young Baek} \email{simply@postech.ac.kr}
\affiliation{Department of Physics, Pohang University of Science and Technology (POSTECH), Pohang, 790-784, Korea}

\author{Yong Wook Cheong}
\affiliation{Department of Physics Education, Seoul National University, Seoul, 151-742, Korea}

\author{Yoon-Ho Kim}
\email{yoonho@postech.ac.kr}
\affiliation{Department of Physics, Pohang University of Science and Technology (POSTECH), Pohang, 790-784, Korea}

\date[Revised:]{ \today}

%%%%%%%%%%%%%%%%%%%%%%%%%%
\begin{abstract}
We propose and demonstrate a linear optical device which deterministically performs optimal quantum measurement or minimum disturbance measurement on a single-photon polarization qubit with the help of an ancillary path qubit introduced to the same photon. We show theoretically and experimentally that this device satisfies the minimum disturbance measurement condition by investigating the relation between the information gain (estimation fidelity) and the state disturbance due to measurement (operation fidelity). Our implementation of minimum disturbance measurement is postselection-free in the sense that all detection events are counted toward evaluation of the estimation fidelity and the operation fidelity, i.e., there is no need for coincidence postselection of the detection events.
\end{abstract}
%%%%%%%%%%%%%%%%%%%%%%%

\pacs{03.67.-a,03.65.Wj, 42.50.Dv, 42.50.-p}

\maketitle

%%% Introduction %%%

A measurement is a process by which we learn about the observed
system by interacting it with a measuring apparatus. The role of the
measurement process is one of the most unique features that
distinguish quantum physics from classical physics \cite{peres}. One
of the fundamental aspects of the quantum measurement process is
that the quantum state of the observed system is unavoidably altered
by the measurement process itself which may be direct, as manifested
in the Heisenberg's uncertainty principle \cite{heisenberg}, or
indirect, as demonstrated in the quantum eraser-type test on the
complementarity \cite{scully82,kim00}.

Quantitative study of on the relation between the information gain
by a measurement and the measurement-induced state disturbance is, obviously, a
relevant and an important issue in quantum physics and quantum
information \cite{note}. In particular, how to achieve the optimal
quantum measurement process, in which the information gain is
maximal while the measurement-induced state disturbance is minimal,
is an important fundamental as well as a practically relevant
problem in quantum communication \cite{massar,bruss,fuchs}.

Recently, the trade-off relation between the information gain and
the state disturbance was derived in the context of a finite
$d$-dimensional quantum system by quantifying the information gain
as the estimation fidelity and the state disturbance as the
operation fidelity \cite{banaszek01}. The measurement protocol
which saturates the trade-off relation is known as the minimum
disturbance measurement (MDM). Experimental demonstrations of MDM to date, however, have been rather limited.
In Ref.~\cite{sci06}, a MDM protocol was implemented for a
single-photon polarization qubit with an ancilla single-photon
polarization qubit by using classical active feed-forward and a
linear optical nondeterministic quantum logic operation based on
coincidence postselection of detection events \cite{todd}. The scheme, therefore, is probabilistic in principle and post-selection of the final detection events is necessary. In
Ref.~\cite{andersen}, a MDM protocol was demonstrated for an
infinite dimensional Gaussian state by using
linear optics, amplitude and phase modulators, and homodyne
detection. This scheme, therefore, applies to a coherent state but not to a qubit. 

In this Letter, we propose and demonstrate a novel
postselection-free linear optical MDM device which deterministically performs optimal quantum measurement of the polarization qubit of a single-photon. The ancilla
qubit which interacts with the polarization qubit is the path qubit
introduced to the same photon, i.e., our device makes use of the
single-photon polarization-path two-qubit state \cite{kim}. We first
show theoretically that this device performs minimum disturbance
measurement by investigating the trade-off relation between the
estimation fidelity and the operation fidelity. We then demonstrate
that the experimental estimation and operation fidelities indeed
closely follow the theoretical bound for MDM. Our MDM is postselection-free in the sense that all detection events are counted toward the evaluation of the operation and estimation fidelities. To the best of our knowledge, postselection-free MDM for a qubit has not been reported to date. 

%%% MDM Proposal %%%

%%%%%%%%%%%%%%%%%%
\begin{figure}[b]
\centering
\includegraphics[width=2.5in]{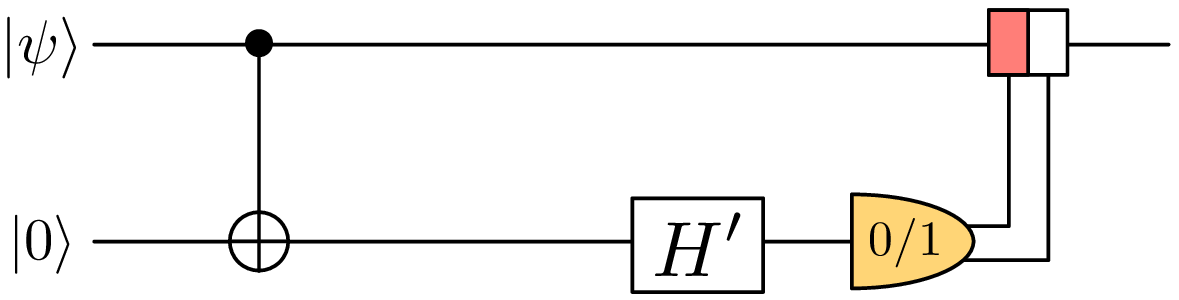}
\caption{Quantum circuit for the proposed MDM on the system qubit
$|\psi\rangle$. $H'$ is a Hadamard-like transformation on the
ancilla qubit and controls the amount of information gain on the
system qubit via measurement on the ancilla qubit.}\label{fig1}
\end{figure}
%%%%%%%%%%%%%%%%%

The quantum circuit for the proposed MDM scheme is shown in
Fig.~\ref{fig1}. The system qubit, which is to be measured, is prepared in
an arbitrary quantum state
$|\psi\rangle_s=\alpha|0\rangle_s+\beta|1\rangle_s$ with
$|\alpha|^{2}+|\beta|^{2}=1$. The ancilla qubit, initialized in the state $|0\rangle_a$, is introduced to make optimal measurement on the system qubit by interacting with it. The controlled-not (CNOT)
operation between the system and the ancilla qubits transforms the initial
two-qubit state $|\psi\rangle_s \otimes |0\rangle_a$ into $\alpha|0\rangle_s|0\rangle_a +
\beta|1\rangle_s |1\rangle_a$, i.e., the system and the ancilla qubits are now entangled. The ancilla qubit then undergoes to
the Hadamard-like unitary transformation $H^{\prime}=
\left(\begin{array}{cc}t & r \\r & -t
\\ \end{array} \right)$. Here $t$ and $r$ satisfy the normalization
condition $|t|^{2}+|r|^{2}=1$ and we assume $|t| \geq |r|$ without
loss of generality.

After the $H'$ operation, the joint state of the
system and the ancilla qubits is given as
%%%%%%%%%%%%%%%%%%%%%%%
\begin{equation}\label{eq1}
(\alpha t |0\rangle_s+\beta r |1\rangle_s)|0\rangle_a + (\alpha r
|0\rangle_s-\beta t |1\rangle_s)|1\rangle_a.
\end{equation}
%%%%%%%%%%%%%%%%%%%%%
First, consider the case of $t=r$ which is equivalent to randomly guessing the state of the system qubit. If the ancilla measurement outcome is 0, the state of the system qubit is unchanged but if the ancilla measurement outcome is 1, it is necessary to apply the feed-forward $\sigma_z$ operation to the system qubit, i.e., $|1\rangle_s \rightarrow -|1\rangle_s$, to recover the original quantum state. Next, consider the case of $t \neq r$. More information about the system qubit can be obtained at the expense of increased state disturbance. The quantum circuit in Fig.~\ref{fig1}, therefore, allows us to vary the ``strength'' of the measurement on the system qubit with variables $t$ and $r$, hence varying the amount of information we can gain about the system qubit from the measurement outcomes of the ancilla qubit.

To see if the quantum circuit in Fig.~\ref{fig1} satisfy
the minimum disturbance condition in Ref.~\cite{banaszek01}, it is necessary to investigate the relation between the estimation fidelity $G$ and the operation fidelity $F$. Consider the state in eq.~(\ref{eq1}). The probabilities of ancilla measurement outcomes 0 and 1 are calculated to be $P_{0}=|\alpha|^{2}t^{2}+|\beta|^{2}r^{2}$ and $P_{1}=|\alpha|^{2}r^{2}+|\beta|^{2}t^{2}$, respectively. The state of the system qubit is then estimated to be
$\rho_G=P_{0}|0\rangle_{s}{}_s\langle0|+P_{1}|1\rangle_{s}{}_s\langle1|$,
i.e., when the measurement outcome of the ancilla qubit is $i$ ($i=0$ or 1), the
system qubit is guessed to be in the state $|i\rangle_{s}$ with probability of
$P_{i}$. The overlap between the the inferred (guessed) state of the
system qubit $\rho_G$ and the original state of the system qubit
$|\psi\rangle_s$ is defined to be the estimation fidelity $G_{\psi}={}_s\langle \psi
| \rho_G |\psi\rangle_s$ and is dependent on the measurement strength controlled by $H'$. If
we consider all possible pure states on the Bloch sphere
as the system qubit, the average estimation fidelity is calculated to be
%%%%%%%%%%%%%%%%%%
\begin{equation}\label{eq2}
G_{\textrm{avg}}=\int G_{\psi} d \psi = \frac{1}{3}(t^{2}+1).
\end{equation}
%%%%%%%%%%%%%%%%

The state of the system qubit after the feed-forward operation in Fig.~\ref{fig1} can be evaluated by tracing over the Hilbert space of the ancilla qubit on the final system-ancilla two-qubit state and is found to be $\rho_F=|\psi_{0}^{\prime}\rangle_{s}
{}_{s}\langle\psi_{0}^{\prime}|+|\psi_{1}^{\prime}\rangle_{s}
{}_{s}\langle\psi_{1}^{\prime}|$ where
$|\psi_{0}^{\prime}\rangle_s=\alpha t |0\rangle_s+\beta r
|1\rangle_s$ and $|\psi_{1}^{\prime}\rangle_{s}=\alpha r
|0\rangle_s+\beta t |1\rangle_s$. The operation fidelity which quantifies the overlap between the input state $|\psi\rangle_s$ and the output state $\rho_F$ is defined as $F_\psi={}_s\langle\psi|\rho_F|\psi\rangle_s$. As before, by averaging over all possible pure states on the Bloch sphere, the average operation fidelity
becomes
%%%%%%%%%%%%%%%%%%%
\begin{equation}\label{eq3}
F_{\textrm{avg}}=\int F_{\psi} d \psi =\frac{2}{3}(1+tr).
\end{equation}
%%%%%%%%%%%%%%
Note that the measurement strength setting which is equal to
the classical random guess, i.e., $t=r=1/\sqrt{2}$, results in
$F_{\textrm{avg}}=1$. In other words, the system qubit has not been disturbed. Any
stronger measurement, however, will reduce the average operation fidelity $F_{\textrm{avg}}$. By equating eq.~(\ref{eq2}) and eq.~(\ref{eq3}), we arrive at the trade-off relation between $G_{\textrm{avg}}$ and $F_{\textrm{avg}}$
%%%%%%%%%%%%%%%%%%%%%%%
\begin{equation}\label{eq4}
F_{\textrm{avg}}=\frac{2}{3}+\frac{\sqrt{1-(6G_{\textrm{avg}}-3)^2}}{3},
\end{equation}
%%%%%%%%%%%%%%%%%%%%%%%
which in fact is the exact minimum disturbance measurement condition in Ref.~\cite{banaszek01} for a qubit.

%%% MDM Experimental Setup %%%

%%%%%%%%%%%%%%%%%%
\begin{figure}[t]
\centering
\includegraphics[width=3.2in]{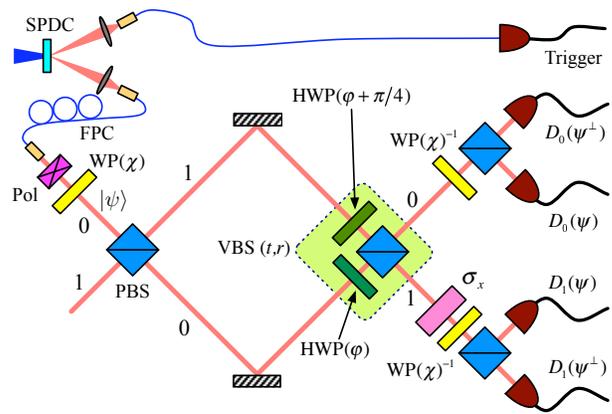}
\caption{Experimental setup. The heralded single-photon state was
used to implement the postselection-free linear optical MDM device
shown in Fig.~\ref{fig1}. A fiber polarization controller (FPC), a
vertical polarizer (Pol), and WP($\chi$) prepare the initial
polarization qubit $|\psi\rangle$. VBS($t,r$) is realized by using a
pair of HWP's and a PBS. }\label{setup}
\end{figure}
%%%%%%%%%%%%%%%%%

The actual experimental setup to realize the quantum circuit in
Fig.~\ref{fig1} is schematically shown in Fig.~\ref{setup}. To
prepare the single-photon state needed to implement the proposed MDM
device, we employ the heralded single-photon source using
spontaneous parametric down-conversion (SPDC) \cite{hong,baek08}.
The SPDC signal-idler photon pair was generated in a 3 mm thick
type-II BBO crystal pumped by a frequency-doubled ultrafast pulses
from a mode-locked Ti:Sapphire laser operating at 780 nm. The pump
beam was focused on the BBO crystal with a lens ($f=300$ mm) and the
780 nm signal-idler photon pair, emitted in the beam-like
configuration at $\pm3.38^\circ$ with respect to the pump beam, was
coupled into single-mode fibers using $\times 10$ objective lenses
located at 650 mm from the crystal \cite{kim03}.

The idler photon was directly coupled to the trigger detector so
that the detection signal can be used to herald the single-photon
state for the signal photon. A fiber polarization controller (FPC)
and a vertical polarizer (Pol) prepare the initial polarization
state of the signal photon at $|V\rangle$. It is then transformed
unitarily to an arbitrary polarization qubit
$|\psi\rangle_s=\alpha|H\rangle_s+\beta|V\rangle_s$ using
WP($\chi$), which consists of a HWP at an angle $\theta_1$ followed
by a QWP at an angle $\theta_2$. (The vertical polarization is
defined to be $0^\circ$.) The ancilla path qubit, which is
initialized at $|0\rangle_a$, is introduced to the same
single-photon state by directing the photon at one (labeled as 0) of
the two input modes of the PBS. The other input spatial mode of the
PBS, labeled 1, is not used. For the prepared initial two-qubit
state $ |\psi\rangle_s\otimes|0\rangle_a$, the CNOT operation
between the polarization qubit and the path qubit is implemented
with the PBS \cite{kim,cerf}.

A balanced Mach-Zehnder interferometer with a variable beam splitter
VBS($t,r$), which consists of a PBS and a HWP in each of the input
modes of the PBS, then implements the Hadamard-like unitary
transformation $H'$ on the path qubit, i.e., creating arbitrary
quantum superposition between the two orthonormal basis vectors
$|0\rangle_a$ and $|1\rangle_a$ of the ancilla path qubit. The angle
of HWP in path 0 is set at $\varphi$ and the corresponding angle of
the HWP in the path 1 is set at $\varphi+\pi/4$. The transmission
and the reflection coefficients of the HWP-PBS system is then given
as $t=\cos2\varphi$ and $r=\sin2\varphi$. Thus, by varying the HWP
angle $\varphi$ we can control the strength of measurement on the
system (polarization) qubit. Note that the initial PBS, which implements the CNOT
operation between the polarization qubit and the path qubit, and the
second PBS, which implements $H'$ operation on the path qubit, form
a balanced Mach-Zehnder interferometer.

A single-photon emerging at the output mode 0 or 1 of VBS($t,r$)
would then mean measurement of the path qubit with the outcome 0 or
1, respectively. We guess (or infer) that, if the measurement
outcome of the path qubit is 0 or 1, the input polarization qubit
must have been in the $|H\rangle$ or $|V\rangle$ state,
respectively, with a certain fidelity. For a single-photon emerging
at the output mode 0 of  VBS($t,r$), i.e., the outcome of
measurement on the ancilla path qubit is $|0\rangle$, the state of
the system qubit (the polarization state) must be $\alpha t
|H\rangle+\beta r |V\rangle$. Similarly, for a single-photon
emerging at the output mode 1 of the HWP-PBS system, the
polarization state must be $\beta t|H\rangle+\alpha r|V\rangle$. We
thus apply the conditional feed-forward operation $\sigma_x$ in the output mode 1 to
flip the polarization state so that
$|H\rangle\leftrightarrow|V\rangle$ and this operation was
implemented with a HWP set at $45^\circ$. (Note that $\sigma_z$ was needed for the protocol described in Fig.~\ref{fig1}.)

Finally, to perform state analysis on the output system qubit (the
polarization state), we first apply the inverse polarization
transformation WP($\chi$)$^{-1}$ by using a QWP at an angle
$\theta_{2}+\pi/2$ followed by a HWP at an angle $\theta_{1}$, in
each output mode of the Mach-Zehnder interferometer. The state of
the output system qubit was then analyzed by using a PBS followed by
two detectors, $D_{i}(\psi)$ and $D_{i}(\psi^\perp)$, at the output
mode $i$ of the Mach-Zehnder interferometer. Here,
$\langle\psi|\psi^\perp\rangle=0$. Keep in mind that the subscript
$i$ ($i=0$ or 1) denotes the measurement outcome of the ancilla path
qubit. Since we are dealing with the heralded single-photon source,
we record the coincidence count between the trigger detector and one
of the four detectors $D_{i}(\psi)$ and $D_{i}(\psi^\perp)$. Note that the coincidence measurement needed for heralded single-photon state has nothing to do with to the postselection-free feature of the present MDM protocol.

%%% Procedure and Data Analysis Method %%%
%%%%%%%%%%%%%%%%%%%%%%%%%%%%%%
\begin{table}[b]
\caption{An experimental data set for $\varphi=22.5^\circ$. Channel efficiency corrected coincidence counts (Hz), averaged over four independent measurements, are shown for six different input states. This data set is used to evaluate the $G_{\textrm{avg}}$ and $F_{\textrm{avg}}$. This measurement is then repeated three times.}
\begin{ruledtabular}
\begin{tabular}{c c c c c}
  $\psi$ & $N_{0}(\psi)$ & $N_{0}(\psi^{\perp})$ & $N_{1}(\psi)$ & $N_{1}(\psi^{\perp})$\\[0.5ex]\hline
  $|H\rangle$ & 71.72 & 0.48 & 73.73 & 0.40 \\
  $|V\rangle$ & 82.97 & 0.26 & 77.08 & 0.11 \\
  $|D\rangle$ & 77.69 & 0.67 & 78.07 & 2.31 \\
  $|A\rangle$ & 80.73 & 1.67 & 73.51 & 2.97 \\
  $|R\rangle$ & 78.07 & 1.33 & 74.16 & 4.95 \\
  $|L\rangle$ & 83.02 & 2.00 & 75.46 & 1.65 \\
\end{tabular}\end{ruledtabular}\label{table1}
\end{table}
%%%%%%%%%%%%%%%%%%%%%%%%%%%%%%%%

The experimental estimation fidelity $G_{\textrm{avg}}$ and the
operation fidelity $F_{\textrm{avg}}$ are obtained from the
detection events as follows. First, the measurement strength on the
ancilla path qubit was set by choosing the HWP angle $\varphi$
between $0^\circ$ and $22.5^\circ$. When $\varphi=22.5^\circ$, $H'$
becomes the usual Hadamard operation on the path qubit since
$t=r=1/\sqrt{2}$ and this corresponds to the weakest measurement or the classical random guess. The strongest
measurement setting is $\varphi=0^\circ$.

Second, for an input system qubit (the polarization state)
$|\psi\rangle_s$, we measure the coincidence counts between the
trigger detector and one of the four state analyzing detectors
$D_{i}(\psi)$ and $D_{i}(\psi^\perp)$ while keeping the phase
difference between the two arms of the Mach-Zehnder interferometer
at 0, modulo $2\pi$. Since the multi-mode fiber coupled detectors
have slightly different coupling efficiencies, the raw coincidence
counts are then corrected for the measured coupling efficiencies,
resulting $N_{i}(\psi)$ and $N_{i}(\psi^{\perp})$. The normalized
count is defined as $n_{i}(\psi)=N_{i}(\psi)/N_{tot}$, where
$N_{tot}= N_{0}(\psi)+ N_{0}(\psi^{\perp})+N_{1}(\psi)+
N_{1}(\psi^{\perp})$, and similarly for $n_{i}(\psi^{\perp})$.

Third, the state dependent operation fidelity $F_\psi$ is then
evaluated as
$F_\psi=\langle\psi|\rho_F|\psi\rangle=n_{0}(\psi)+n_{1}(\psi)$. The
state dependent estimation fidelity is evaluated as
$G_{\psi}=\langle\psi|\rho_G|\psi\rangle=P_{0}|\langle
H|\psi\rangle|^2+P_{1}|\langle V|\psi\rangle|^2$, where the
probability of measurement outcome of $i$ is given as
$P_{i}=n_{i}(\psi)+n_{i}(\psi^{\perp})$. The state-dependent
$G_\psi$ and $F_\psi$ are then obtained for six different input
system qubits, i.e., $|H\rangle$, $|V\rangle$,
$|L\rangle=(|H\rangle+i|V\rangle)/\sqrt{2}$,
$|R\rangle=(|H\rangle-i|V\rangle)/\sqrt{2}$,
$|D\rangle=(|H\rangle+|V\rangle)/\sqrt{2}$,
$|A\rangle=(|H\rangle-|V\rangle)/\sqrt{2}$ \cite{sci06}. The state-averaged estimation and operation fidelities $G_{\textrm{avg}}$ and $F_{\textrm{avg}}$ are then obtained by averaging $G_\psi$ and $F_\psi$ values, respectively, for the six input states. Finally, the whole procedure
was repeated for 10 different settings of the measurement strength,
i.e., 10 different angle settings of $\varphi$.

%%%%%%%%%%%%%%%%%%
\begin{figure}[t]
\centering
\includegraphics[width=3in]{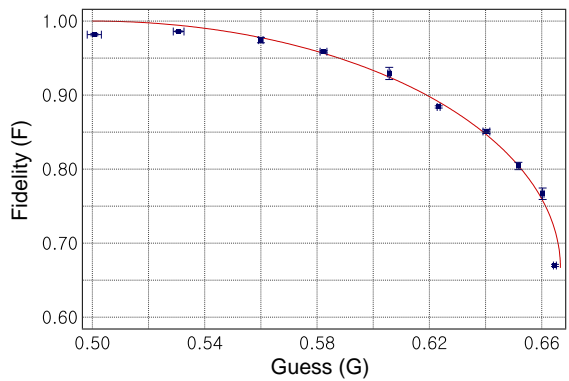}
\caption{Experimental data. Solid line shows the theoretical optimal trade-off relation between the information gain and the state disturbance due to measurement. The experimental estimation fidelity $G_{\textrm{avg}}$ and the operation fidelity $F_{\textrm{avg}}$ are shown as solid squares.
}\label{expdata}
\end{figure}
%%%%%%%%%%%%%%%%

A sample of the experimental data is shown in Table  \ref{table1}. The setting for the measurement, $\varphi=22.5^\circ$, corresponds to the classical random guess, i.e., the theoretical estimation fidelity $G=0.50$. The numbers correspond to channel efficiency corrected coincidence counts (Hz), averaged over four independent measurements, for six different input states. We perform three sets of measurements like in Table \ref{table1} to experimentally evaluate the estimation fidelity $G_{\textrm{avg}}$ and the operation fidelity $F_{\textrm{avg}}$ for each of the measurement setting.

The final experimental data are shown in Fig.\ref{expdata}. The
solid line shows the theoretical optimal trade-off relation of
eq.~(\ref{eq4}) between the information gain (or classical guess;
estimation fidelity) and the state disturbance due to measurement
(operation fidelity). The experimental data are shown in solid
squares and they summarize the results of three experimental runs.
It is clear that the experimentally obtained $G_{\textrm{avg}}$ and
$F_{\textrm{avg}}$ closely follow the MDM condition in
eq.~(\ref{eq4}). Slightly less than ideal operation fidelity,
i.e., $F < 1$, when the estimation fidelity $G = 0.5$ is due to the
imperfect alignment of the Mach-Zehnder interferometer, imperfect
optics, and small efficiency differences between the detectors. (To
reach $F=1$, the Mach-Zehnder interferometer would have to exhibit
100\% interference visibility.)

%%% Conclusion/Summary %%%

In summary, we have proposed a novel minimum disturbance measurement
protocol and the corresponding postselection-free linear optical
scheme to implement the protocol using the single-photon
polarization-path two-qubit state. We have also demonstrated the
proposed scheme using the heralded single-photon source from
spontaneous parametric down-conversion, showing good agreement with
the theoretical $G$-$F$ bound for the optimal quantum measurement. Finally, it is interesting to note that the present protocol and scheme could be expanded to explore optimal quantum measurement bound for high-dimensional quantum states or qudits by using, for example, multi-path interferometric geometries to realize high-dimensional path qudits \cite{weihs,thew}.

This work was supported, in part, by the Korea Science and Engineering Foundation (R01-2006-000-10354-0),  the Korea Research Foundation (KRF-2005-015-C00116 and R08-2004-000-10018-0), and by the Ministry of Commerce, Industry and Energy of Korea through the Industrial Technology Infrastructure Building Program.

%%%%%%%%%%%%%%%%%%%%%%%%%%%%%%%%%%%%%%

\end{document}